\documentclass[12pt]{iopart}
\pdfoutput=1
\expandafter\let\csname equation*\endcsname\relax
\expandafter\let\csname endequation*\endcsname\relax
\usepackage{amsmath}
\usepackage{appendix}
\usepackage{graphicx}
\usepackage{iopams} 
\usepackage{epstopdf}
\usepackage{color} 
\usepackage{float}
\usepackage{times}   

\newcommand{\rr}{{\bf r}}
\newcommand{\la}{\langle}
\newcommand{\ra}{\rangle}
\newcommand{\beq} {\begin{equation}}
\newcommand{\beg} {\begin{equation}}
\newcommand{\eeq} {\end{equation}}
\newcommand{\Phihat}{{\hat \Phi}}

\newcommand{\Phiop}{{\hat \Phi}}
\newcommand{\Ohat}{{\hat O}}

\newcommand{\rhohat}{{\hat \rho}}
\newcommand{\corr}{{\rm corr}}
\newcommand{\Oop}{{\hat O}}
\newcommand{\RR}{{\bf R}}

\newcommand{\av}{{\rm av}}
\newcommand{\maxx}{{\rm max}}
\newcommand{\diff}{{\rm diff}}

\makeatletter

\@ifundefined{textcolor}{}
{%
 \definecolor{BLACK}{gray}{0}
 \definecolor{WHITE}{gray}{1}
 \definecolor{RED}{rgb}{1,0,0}
 \definecolor{GREEN}{rgb}{0,1,0}
 \definecolor{BLUE}{rgb}{0,0,1}
 \definecolor{CYAN}{cmyk}{1,0,0,0}
 \definecolor{MAGENTA}{cmyk}{0,1,0,0}
 \definecolor{YELLOW}{cmyk}{0,0,1,0}
}

\@ifundefined{date}{}{\date{}}
\AtBeginDocument{
  
}
\makeatother

\begin{document}

\title[Density-matrix based numerical methods for discovering order and correlations]{Density-matrix based numerical methods for discovering order and correlations in interacting systems}
\author{Christopher L. Henley}
\address{Laboratory of Atomic And Solid State Physics, Cornell University, Ithaca, NY 14853, USA}
\author{Hitesh J. Changlani}
\address{Department of Physics, University of Illinois at Urbana-Champaign, Urbana, IL 61801, USA}
\date{\today}

\begin{abstract}
We review recently introduced numerical methods for the 
unbiased detection of the order parameter and/or dominant correlations, 
in many-body interacting systems, by using reduced density matrices. 
Most of the paper is devoted to the ``quasi-degenerate density matrix'' (QDDM) 
which is rooted in Anderson's observation that the degenerate 
symmetry-broken states valid in the thermodynamic limit, are 
manifested in finite systems as a set of low-energy ``quasi-degenerate''
states (in addition to the ground state). This method, its original form due to Furukawa et al.
[Phys. Rev. Lett. 96, 047211 (2006)], is given a number of improvements here, 
above all the extension from two-fold symmetry breaking to arbitrary cases. 
This is applied to two test cases (1) interacting spinless hardcore bosons on the triangular lattice 
and (2) a spin-1/2 antiferromagnetic system at the percolation threshold. 
In addition, we survey a different method called the ``correlation density matrix'',  
which detects (possibly long-range) correlations only from the ground state, but 
using the reduced density matrix from a cluster consisting of two spatially separated regions.
\end{abstract}

\maketitle
\section{Introduction}
\label{sec:Introduction}
The ground state of quantum many-body strongly correlated electron, spin or bosonic systems, 
is characterized by a variety of order parameters and/or correlations 
(such as charge density waves, spirals, superconductivity and antiferromagnetism). 
Once the primary objective of finding out if the system is gapped or gapless is achieved 
(i.e. the presence or absence of long range order is established), it is 
important to understand {\it  what kind} of order (or lack of it) exists 
in the system. Even though ``more is different'' and every system is interesting in its own right, 
there are organizational principles that can be uncovered by defining and studying 
metrics which make almost no (or minimal) assumptions of the underlying physics of the system 
and do not rely on human intuition. This is especially important when the 
order is exotic and has not been thought of previously. 
Historically, understanding the order parameter of the system has been an art of sorts: 
one of the main objectives of this Article will be to suggest ways 
in which this entire procedure can be more or less automated.

Given the immense difficulty of the many-body problem and the absence of controlled analytic approximations, 
utilizing numerical methods has become a standard way for understanding quantum systems. 
In particular, exact or highly accurate numerical methods such as 
exact diagonalization (ED), density matrix renormalization group (DMRG)~\cite{White_PRL,Schollwock}
and quantum Monte Carlo (QMC)~\cite{Nightingale_Umrigar,GFMC,Sandvik_QMC,AFQMC} 
have become indispensable tools for 
researchers in this field. It is imperative that we maximize the potential of these methods 
and develop the proper tools to \emph{measure} physical properties of a wide variety of systems.   

In recent years, there has been a cross fertilization of ideas between condensed matter 
and quantum information as a result of which a new language 
for describing strongly correlated systems has emerged. In this language, the 
mathematical object of utmost importance is the quantum mechanical reduced density matrix 
for a given spatial region (other generalizations, for example to momentum space~\cite{Hughes_momentum}, 
have also been explored in the literature). For the present purpose, the term 
"reduced density matrix" $\rhohat_A$ for a spatial cluster $A$, with a local Hilbert space of $D$ states,
is defined  by,
\beq
	\rhohat_A = \text{Tr}_{\bar A} \Big( |\psi_{GS} \ra \la \psi_{GS}| \Big) 	
	\label{eq:rho_A_define}
\eeq
where $|\psi_{GS} \ra$ is the ground state wavefunction of the entire system 
and where ${\rm Tr}_{\bar A}$  means trace over everything but cluster $A$ 
i.e. tracing out the ``environment''. 
Knowledge of the reduced density matrix (henceforth abbreviated as RDM) 
serves a very important purpose: its $D$ eigenvalues 
reflect the probability of the cluster to exist in the corresponding eigenvector, 
just like the thermal density matrix captures the knowledge of the energies 
and microstates of a classical system in contact with a thermal bath. 

While RDMs are central for all the methods covered here, for most of the paper, 
there is a second key ingredient: where the exact eigenstates includes a distinct family of 
low-lying levels which are called {\it quasi-degenerate}
(abbreviated as QD).  We will present a path~(proposed initially by Furukawa et al.~\cite{furukawa}) that 
combines these two ingredients to discovering in an {\it unbiased} fashion what the order parameter is:
we call this promising method the 
{\it quasi-degenerate density matrix} (QDDM) method. 
In our version of the method (and implicitly in Ref.~\cite{furukawa}),
a key step is extending the RDM construction so as to 
mix two different QD states, whence the name ``quasi-degenerate 
density matrix'' (formally defined in Sec.~\ref{sec:QDDM}.)

In the rest of this paper, we first (in Sec.~\ref{sec:FMO}) 
review the original paper of Furukawa, Misguich and Oshikawa~\cite{furukawa}, 
which was limited to two-fold symmetry breakings with two quasi-degenerate states. 
We introduce several improvements that foreshow the extension 
of this method (in Sec~\ref{sec:QDDM}) to an arbitrary number of QD states. 
The method is then tested by computations for two example 
systems: first, in Sec.~\ref{sec:tV}, order parameters for three-fold 
density-wave order on a triangular lattice at 1/3 filling. 
Then in Sec.~\ref{sec:perc}, we characterize emergent $S=1/2$ spins in a quantum 
antiferromagnet at the percolation threshold; 
this shows that the QDDM has applications beyond symmetry breakings 
(the quasi-degeneracy has a different origin in this case.)
In Sec.~\ref{sec:extend}, we survey some as-yet unrealized extensions of the QDDM method. 
In Sec.~\ref{sec:CDM}, we turn to reviewing a distinct method with a similar purpose,
the Correlation Density Matrix (abbreviated as CDM). Finally Sec.~\ref{sec:Conclusion}, 
the conclusion, revisits the most important results and discusses 
the limitations of the various methods.

Before proceeding to the specifics, we want to place this
work against the context of modern ``entanglement'' studies,
a burgeoning industry based on RDMs.
The typical aim of an entanglement calculation (whether numerical or
analytic) is the finite-size scaling of the entanglement 
entropy~\cite{Calabrese_Cardy,Ha10,Is11,Zh11,Lauchli_review,Fradkin_EE,Mcminis_Tubman,Ryu_Takayanagi} 
and/or interpreting the patterns of 
the ``entanglement spectrum''~\cite{Li_Haldane,Lauchli_ES, Calabrese_Lefevre, Bernevig} i.e. the eigenvalues of the RDM. 
The aim is to reliably capture universal, long-wavelength 
properties of topological and/or critical states. For such a calculation the cluster should be as large as is 
practical, e.g. half the total system size (in which case the cluster and environment play equivalent roles). 

Our own interest in RDMs~\cite{cheong-free-fermions,cheong-op-trunc,cheong-2D,cheong-thesis,cheong-CDM,cheong-EGS}
had quite different motivations, (originally inspired by DMRG), eg. 
the discovery of the formula for the free-fermion 
RDM in terms of the Green's function~\cite{cheong-free-fermions}. 
This led to the CDM method ~\cite{cheong-CDM,mue10}, 
which will also be reviewed in Section~\ref{sec:CDM}.
These methods, and the QDDM, use RDMs constructed on the
{\it smallest} clusters that give physical results, typically a few sites.

\section{Basic QDDM: Improving on Furukawa, Misguich and Oshikawa
($M=2$ case)}
\label{sec:FMO}

In this section, we walk through the Quasi-degenerate Density Matrix 
approach in the simplest case, that of two quasi-degenerate eigenstates,
starting from the original paper by Furukawa, Misguich and Oshikawa~\cite{furukawa},
which we henceforth referred to as ``FMO''.
We will explain how one can ``discover''
an order parameter from the RDM, 
in a (nearly) unbiased fashion,
and expose the mathematical/computational apparatus needed.
To make this concrete, one seeks the operator $\Phihat$, 
defined on a cluster of sites $A$, which (acting as a 
perturbation) ``most strongly'' splits the QD states. 

In our retelling of FMO, we take the liberty of using our own notation.
In places, we recast the FMO recipe in a mathematically
equivalent but more illuminating form. 
This will highlight the limitations of the QDDM of 
FMO, which may explain why the method was not taken up and widely applied.
Along the way, we will introduce several major improvements we have made:
a pre-processing requiring no prior assumptions; the use of $L^2$
norms, amenable to eigenvalue analysis; and a better weighting
of the answer.  The greatest extension, the extension to
$M>2$ quasi-degenerate states, is left to the following section.

\subsection{Quasi-degeneracy and order parameters}
\label{sec:Order_parameters_conditions}

Before diving into the methods, we pause in this subsection
to remind the reader of two general concepts for understanding 
how long-range order, in the thermodynamic limit, manifests 
itself in finite systems: quasi-degeneracy 
(mentioned already in the introduction) and order parameter operators.
This section applies also to the general case with $M>2$.

\subsubsection{Quasi-degenerate (QD) states}
\label{sec:QD_states_FMO_sec}

Imagine a situation where we have exact diagonalization results for a system such 
that a subset of $M$ eigenstates $\{ |\psi_m\ra\}$ is {\it quasi-degenerate}
(QD): that is, their eigenenergies are all small compared to those
of all the other states.  In other words, the splittings within
the quasi-degenerate states are small compared to the gap between
them and the other states. This term ``quasi-degenerate'' is
borrowed from the work of Lhuillier and co-workers~\cite{Bernu,Lhuillier_review}, 
who studied such states in the context of antiferromagnetic lattice models.
We have willed the liberty of extending the term to cases in which
the nearly-degeneracy is not due to long-range order in the thermodynamic limit.

One possible origin for quasi-degeneracy is the existence of a spontaneous symmetry breaking. 
As originally understood by Anderson~\cite{anderson-tower}, whenever the associated order parameter 
is not conserved by the Hamiltonian, the actual eigenstates are
{\it not} symmetry-broken states and indeed the order parameter has 
zero expectation in any one of them.  However, the system has a subset of 
quasi-degenerate states, and certain linear combinations of these are 
symmetry-broken states -- which are long-lived, since the quasi-degenerate states differ
only slightly in energy. Equivalently, one can construct a symmetry-broken basis for the QD states; 
the Hamiltonian in this basis has small, off-diagonal terms representing the
small amplitude for tunneling between these states, so the actual
eigenstates are linear combinations with small splittings.
The energy splittings among QD states decrease rapidly with increasing system size.

When calling these states ``quasi-degenerate'', we 
posited that their splitting is to some degree accidental.
In many cases, this splitting depends on details such as 
evenness/oddness of the number of sites, or the aspect ratio 
of the finite system used, and thus may vary in a complicated 
fashion within a family of successively larger systems. 
Therefore, all of our analyses eventually seek to get answers 
invariant with respect to changes of basis among the 
QD states.

\subsubsection{Order parameters and symmetries}

Now we turn to order parameters. Note first that, for any extended system, 
there is an infinite set of valid definitions 
for an order parameter. 
Every one of them transforms the same way under symmetry operations, 
and every one will give the identical answer for the symmetry-broken states. 
Particular order parameters are chosen for convenience -- most often, 
because they are local (e.g. involving one site, or two neighboring sites).

Thus, we are free to adopt whichever definition of order parameter 
leads to a simple solution. This will be controlled by our choice of
the optimization criterion. In view of what was just said about
the freedom in choosing order parameters, we should not worry if
the optimum, according to our trackable criterion, is sub optimal
from the viewpoint of another, more natural, criterion.
On the other hand, 
it is essential that {\it some} optimization is done.
This is to ensure we pick up the fundamental order-parameter
operator, and not a secondary one.
\footnote{
For example, the spin $S_z$ is a fundamental order-parameter 
for an Ising model, whereas $(S_z)^3 - ({\rm const}) S_z$ would
be a secondary operator.}

There {\it is} a major constraint on a proper order-parameter:
it must transform according to a (nontrivial) representation
of the symmetry being broken.  One basic corollary is 
that the order parameter is unbiased: in the order-parameter space,
the expectations from all the symmetry-broken states are 
distributed equally around zero.  Keep in mind that, in general,
we will have multi-component order parameters, and also we have
no idea (a priori) what is the nature of the symmetry.

A particular consequence is that the order parameter,
when averaged over all the QD states, is zero.
\footnote{
This follows since the family of QD states is 
presumed to be related to the symmetry-broken states
by a unitary transformation.}
To translate this notion to the computational context,
we define the {\it average density matrix} 
which is simply the average of the RDM as calculated
from all the QD states:
   \beq
     \rhohat^{\av} \equiv \frac{1}{M} \sum_{m=1}^M {\rm Tr}_{\bar A} \Big( |\psi_m \ra \la \psi_m| \Big) .
   \label{eq:aver-DM}
   \eeq
Note that our desired order-parameter operator $\Phihat$ must satisfy the condition,
    \beq
       \Tr(\rhohat^{\av} \Phihat)=0.
        \label{eq:trace_rho_av}
    \eeq
where the trace here is over all QD states. 

Roughly speaking, the average DM tells us what 
local configurations of the cluster are common to all QD states; 
like a projection matrix, it annihilates any local states that 
occur only in whole-system eigenfunctions with higher energies. 
We are seeking an operator that {\it distinguishes} these states, 
so this is to be found in the subspace {\it orthogonal} to $\rhohat^{\av}$.

\subsubsection{Invariance under change of QD basis}

Finally, we turn to an additional symmetry, related to the
idea of quasi-equivalence.  
The philosophy of the present approach is to view the 
QD states as linear combinations of the symmetry-broken states,
meaning they are idealized as being exactly degenerate.
Therefore, we discard all the information contained in the
splittings among the QD states.  (Note that in other ways 
for using QD states~\cite{Lhuillier_review}, the crucial
clues to the nature of the long-range order holding in
the thermodynamic limit, is precisely the pattern of splittings,
as correlated with lattice and spin symmetries of the states.)

It follows that any construction we use must give identical
answers, {\it independent} of any unitary transformation
of the basis we use for the QD subspace of the Hilbert space,
  \beq
       |\psi'_{m'} \ra \equiv \sum_m U_{m'm} |\psi_m \ra
  \eeq
where $U_{m'm}$ is an $M\times M$ unitary matrix and $|\psi'_{m'} \ra$ is a transformed QD state.
Evidently, the average density matrix (Eq.~\eqref{eq:aver-DM})
satisfies this condition, being invariant under such transformations.

Now consider the implications for the reduced density matrix itself.
If we let $\rhohat^{'m'}$ denote the RDM constructed from 
$|\psi'_{m'}\ra $, we see
    \beq
         {\rhohat}^{'m'} = \sum_{mn} (U_{m'm} U^{*}_{m'n}) \rhohat^{mn} 
    \eeq
involving a generalization of the density matrix that is {\it off-diagonal}
in that it is constructed from two {\it different} eigenstates $m$ and $n$:
   \beq
   \rhohat^{mn} \equiv {\rm Tr}_{\bar A} \Big(|\psi_m\ra \la \psi_n| \Big).
   \label{eq:QDDM}
   \eeq
When, as here, $\rhohat^{mn}$ is built from the QD states, 
we call it the {\it quasi-degenerate density matrix} (QDDM).
Note that $\rhohat^{mn}$ is implicitly dependent on the choice of cluster $A$;
thus $A$ should be considered as yet another index of $\rhohat^{mn}$:
that has been suppressed, since the cluster $A$ once chosen is 
kept fixed through the analysis.

Off-diagonal density matrices can appear in other contexts, 
involving different choices of the subspace of eigenstates 
entering them. In any case, the off-diagonal density matrix contains information
on any matrix element of any operator $\Oop$ (defined on $A$)
between any two of the eigenstates:
\begin{equation}
\la \psi_m | \Oop | \psi_n\ra = {\rm Tr_{A}}\Big(\rhohat^{mn} \Oop \Big)
\label{eq:expect_wt_QDDM}
\end{equation}
(To verify this, simply insert equation~\eqref{eq:QDDM}
in the right hand side of equation~\eqref{eq:expect_wt_QDDM}, and use the definition of the trace.)

\subsection{Initial symmetry processing of QD states}
\label{sec:Ini_sym_proc}

FMO begin by assuming there are just $M=2$ QD states, $\psi_1$ and $\psi_2$.
The final aim is to find (1) the order-parameter operator $\Phihat$,
and (2) the two linear combinations of $\psi_1$ and $\psi_2$ 
which approximate the symmetry-broken states,
as best as possible in a system of finite-size.
The goals are related in that $\Phihat$ {\it is} the operator that 
splits the linear combinations of $\psi_1$ and $\psi_2$ as much as possible.

In the case of $M=2$ states, we have a $2^2-1 =3$ dimensional 
space of operators 
orthogonal to $\rhohat^{\av}$. 
FMO {\it assume} time-reversal and/or lattice symmetries,
which allow them to find a special basis for the DM states 
$\Psi_1$ and $\Psi_2$.  Then they uniquely define the operator 
$\rhohat^{\diff} = \rhohat_{11}-\rhohat_{22}$. 
(In the presence of either symmetry, the other two QDDM components 
$\rhohat_{12}$ and $\rhohat_{21}$ are in fact zero, making their search for 
an order parameter one dimensional.)

The symmetry assumption unnecessarily violates our principle
of using no assumptions about the order. In fact, one can find the {\it same} special DM basis as 
FMO even in the {\it absence} of symmetries. Roughly speaking, 
the criterion for the optimal basis vectors is to minimize
the operator norms of $\rhohat_{12}$ and $\rhohat_{21}$, 
the off-diagonal parts of the QDDM. 

The above mentioned criterion is a special case of a method 
we will propose for general $M$ (the modified viewpoint) explained in Sec.~3.2. 
Since the details depend on using fused indices and rely on a few more formal definitions, 
we will defer discussions about our method to this later section.

\subsection{Optimization criterion ($M=2$ case)}
\label{sec:FMO-opt-criterion}

FMO chose the optimization criterion, 
   \beq
	\sigma_{\Phi} = \text{max}_{||\Phi||^{\maxx}=1} 
        \left|\Tr(\Phi \rhohat^{\diff} ) \right|
        \label{eq:sigma_phi}
   \eeq
where the trace is over the $M=2$ QD states and the notation $||\Ohat||^{\maxx}$ denotes a kind of operator norm,
   \beq
       ||\Ohat||^{\maxx} \equiv {\rm max}  |\langle v|\Ohat|v \rangle| 
        \label{eq:op_norm}
   \eeq
where $v$ is any normalized state vector.
\footnote{
This $||...||^{\maxx}$ norm is reminiscent of an $L_\infty$ norm,
except that it is properly invariant with respect to changes of basis.}
The solution to this optimization problem~\eqref{eq:sigma_phi} is shown to be
\begin{equation}
	{\Phi} = \sum_{j} |j \rangle \text{sgn}\;\;\lambda_j \langle j|
        \label{eq:FMO_phi}
\end{equation}
where {$\lambda_j$} are the eigenvalues and 
{$|j\rangle$} are the corresponding eigenvectors of the 
operator $\rho^{\diff}$, 
and sgn${\lambda_j}$ is the sign of $\lambda_j$.
This form suffers from the deficiency that it gives an 
equal weight to the leading eigenvector, 
as it does to those with negligible eigenvalue.  
That will be pernicious in numerical applications, 
when the smaller $\lambda_j$'s are dominated by roundoff errors. 

Instead, a much more numerically stable form is obtained, 
if we take a criterion using $L^2$ norms:
   \beq
	\sigma_{\Phi} = \text{max}_{||\Phi||^2=1} 
        \left|\Tr(\Phi \rhohat^{\diff} ) \right|^2
        \label{eq:L2_norm}
   \eeq
where, 
    \beq
        ||\Phi||^2 \equiv \Tr(\Phi^{*} \Phi) = \sum_{mn} |\la \psi_m|\Phi | \psi_n \ra |^2
        \label{eq:std_op_norm}
    \eeq
is the standard operator norm.

As in the FMO method, diagonalizing $\rho^{\diff}$ (and after imposing the normalization condition on $\Phi$), 
one obtains the solution, which is,
\begin{equation}
	{\Phi} = \frac{1}{\sqrt{2}}\sum_{j} \lambda_j |j \rangle \langle j| = \frac{1}{\sqrt{2}} \rhohat^{\diff}.
\end{equation}
In this form, the subdominant eigenvectors are suppressed.

\subsection{Choice of the cluster size}
\label{sec:FMO-chose-cluster}

The cluster $A$ should generally be the smallest one that will 
capture the order parameter. 
This means that the resulting recipe for the order parameter 
is {\it local}, which as mentioned above 
(Sec. \ref{sec:Order_parameters_conditions}) is usually desirable.
FMO illustrated this choice of cluster with the contrasting 
examples of N\'eel order in a spin chain (requiring one site) 
with dimer order (requiring two). But one may choose a larger cluster that may be more 
symmetric with respect to the order parameter. 

This all highlights that the cluster choice is the weak 
link in our aim to have an unbiased method: 
if we do not know the order parameter {\it a priori}, 
we cannot be sure what cluster $A$ suffices to capture it. 
We do not have a definitive answer, since we already had an 
expectation when we approached the example cases 
(Sections \ref{sec:tV} and \ref{sec:perc} in this paper). 
We believe this will show up in the final stage of 
analyzing the resulting operators. 
If the cluster is too small, 
we will probably find that the operators obtained are insufficient 
to fully split the QD manifold of states.

The same cluster should be used in QDDM calculations using
different system sizes. The resulting order parameters should
be exactly the same operators, in principle. 
Deviations from this behavior give a measure of finite size effects 
and whether the results are adequately converged.

The QDDM of Ref.~\cite{furukawa} was applied not only to systems
with an order parameter, but also to those with {\it topological order}.  
The analog of an order parameter must be a string operator involving
a set of sites forming a topologically nontrivial loop. 
(Here is one situation that calls for an extended cluster $A$). 
 FMO applied this to the case of ($T_2$) topological order. 
One could combine our generalization of our $M>2$ method 
(in Sec.~\ref{sec:QDDM}) to topological order, and capture 
more general forms of it.

\section{Extended QDDM: Recipe for many quasi-degenerate states}
\label{sec:QDDM}

Now we turn to the harder task of extending the ideas of Sec.~\ref{sec:FMO} 
to general, multi-component order parameters that go along with $M>2$ 
quasi-degenerate states. 

Trying to imitate the path of Sec.~\ref{sec:FMO}, we immediately 
run across two questions. 
The first and easier one is, which cost function to choose? 
The second question is how to generalize  the ``preprocessing'' described in 
Sec~\ref{sec:Ini_sym_proc}. As noted already, 
the symmetric combination of 
density matices $\rhohat^\av$ cannot split the QD states, so we 
must certainly work in the orthogonal subspace, which evidently has
dimension $M^2-1$ in general.
In the FMO treatment of the $M=2$ case, that 3-dimensional subspace
was reduced to a one dimensional subspace called $\rhohat^\diff$.
And in fact, this was basically the final answer, 
since the order parameter was also one dimensional.
For $M>2$, we face the new task of somehow finding the right
subset of that $M^2-1$-dimensional ``difference''  space:
the sought-for order parameter space has a lower dimension.

\subsection{Fusing indices and operator singular value decomposition}
\label{sec:Q=3-optimal}
As in Sec.~\ref{sec:FMO} we were looking for the (normalized) operator $\Phihat$,
defined within cluster $A$, that has the ``maximum possible'' expectation, 
if we are allowed to make any (normalized) linear 
combination of the QD states $\{ |\psi_m\ra \}$. 
Following the philosophy of using $L^2$ normals, 
we choose to maximize
   \beq
    \sigma^2_\Phi
     = {\sum_{mn}}^{'} \big|\la\psi_m|\Phihat|\psi_n\ra\big|^2
   \label{eq:sigma_2}
   \eeq
In other words, our criterion is to maximize the 
{\it variance} of the order parameter matrix elements
rather than to find an operator that maximizes its 
magnitude acting only on a single symmetry broken state.
(A practical drawback of the latter choice is 
that the action of the order parameter on the sub-maximally 
split states would be underdetermined.)
Instead, all states (and all possible operators) must be treated on an equal footing.

Note that the "$\sum'$" symbol in Eq.~\eqref{eq:sigma_2}
signifies we are using a QDDM with $\rhohat^\av$ projected out,
i.e. part of the ``difference'' subspace  having rank $M^2-1$.
Then if an order parameter operator $\Phi$ had any component
of $\rhohat^\av$, it would contribute nothing to our objective
function \eqref{eq:sigma_2}.  Therefore, with this criterion
all components of the order parameter must indeed be constructed
from the ``difference'' subspace.
From here on, all our work is within that subspace.

We now use Eq.~\eqref{eq:expect_wt_QDDM}, to rewrite Eq.~\eqref{eq:sigma_2} as,
     \beq
    \sigma^2_{\Phi}
= {\sum_{mn}}' \text{Tr}
  \left( \rho^{mn} \Phi \right) \text{Tr} \left( \rho^{mn} \Phi \right)^{*},
     \label{eq:variance_QDDM2}
     \eeq

Then, each factor in our objective function~\eqref{eq:variance_QDDM2} can be re-expanded:
\begin{eqnarray}
     \text{Tr} \left( \rho^{mn} \Phi \right) = \sum_{aa'} \langle a | \rho^{mn}|a'\rangle \langle a'|\Phi| a \rangle
                                             \equiv \sum_{aa'} \rho^{mn}_{aa'} \Phi_{a'a} \label{eq:QDDM_aa'}
\end{eqnarray}
where we have used $ a (a')$ as the index for the basis state for cluster 
$A$ and have defined the notation $\rho^{mn}_{aa'} \equiv  \langle a | \rho^{mn}|a'\rangle $. 

To make progress, we can handle Eq.~\eqref{eq:variance_QDDM2} using the tricks
of grouping indices and defining fused indices. 
More specifically, we think of $mn$ as one fused index (denoting a compact index for a pair of QD states) 
and $aa'$ as another (denoting a compact index for a pair of states of the cluster Hilbert space), so that the 
matrix elements of the QDDM are $\rho_{mn,aa'} \equiv \la a|\rhohat^{mn}|a'\ra$. 
(This very same matrix element $\rho_{mn,aa'}$ has been written out as $\rho^{mn}_{aa'}$ for ease of notation.)
Another way of saying this is that the QDDM is an $M\times M$ array of operators (as was explained in Sec. 2.1.3), 
each of which is a diagonal or off diagonal reduced DM, constructed from 
two (possibly different) eigenstates from the QD subspace. In our language, 
the QDDM is a rectangular matrix of dimension $M^2 \times D^2$, 
where $D$ is the dimension of the Hilbert space of cluster $A$.

We define fused indices for the local Hilbert space indices $aa' \to  \alpha$
and $bb' \to  \beta$, and $mn \to \mu$ for the QD indices:
Then we rewrite equation~\eqref{eq:QDDM_aa'} 
using the fused indices; we make the replacements 
$\sum_{aa'} \rho^{mn}_{aa'} \Phi_{a'a} \to \sum_\alpha \rho^{\mu}_\alpha \Phi^{*}_\alpha$,
where we have used the Hermiticity of $\Phi$ i.e. $\Phi_{a'a} = \Phi_{aa'}^{*}$
Hence Eq.~\eqref{eq:variance_QDDM2} becomes,
    \beq
         \sigma^2_\Phi = {\sum_{\mu\alpha\beta}}^{'}
           \rho^{\mu}_{\alpha} {\rho^{\mu}_{\beta}}^{*} \Phi_\alpha^{*} \Phi_\beta
    \label{eq:QD-quad-objective}
    \eeq
Note that with the adoption of fused indices, the $D \times D$ matrix $\Phi_{a'a}$ is re-framed as
a vector $\Phi_{\alpha}$ of size $D^2$.

To put the objective function in equation~\eqref{eq:QD-quad-objective} into 
a more illuminating form, define the contracted matrix
    \beq
        C_{\alpha\beta} \equiv
           {\sum_{\mu}}^{'} \rho^{\mu}_{\alpha} {\rho^{\mu}_{\beta}}^{*} 
        \label{eq:contracted_matrix}
    \eeq
which can easily be verified to be Hermitian. At this point we re-emphasize the importance of the $\sum'$, 
it is crucial when computing the contracted matrix. 

Finally, our objective function becomes,
    \beq
    \sigma^2_\Phi
          = \sum_{\alpha\beta} C_{\alpha\beta} \Phi_\alpha^{*} \Phi_\beta
        \label{eq:sigma_phi_QDDM}
    \eeq
Note that this is nothing but a vector-matrix-vector contraction.

The next (and key) step is to diagonalize $C_{\alpha\beta}$, 
which is an example of the advertised 
``operator singular value decomposition''. 
In the diagonal basis, the $C_{\alpha\beta}$ is replaced by
a transformed matrix ${\tilde C}_{jk} = |\lambda_j|^2 \delta_{jk}$,
where $|\lambda_j|^2$ are the eigenvalues. We get
   \beq
    \sigma^2_\Phi  = \sum_j |\lambda _j|^2 {\tilde \Phi_{j}} {\tilde \Phi_{j}}^*,
    \label{eq:final-sum}
   \eeq
where $\tilde \Phi_{j}$ are the eigenvectors of the contracted matrix. 
Note that the coefficients of eigenvector $\Phi_{j}$ in the $\alpha$ basis can be denoted as $\Phi^{(j)}_{\alpha}$.
Also, realize that we are allowed the call the eigenvalues ``$|\lambda_j|^2$'' since
$C_{\alpha\beta}$ is a nonnegative-definite matrix by construction;
a better reason for this notation will emerge in the next subsection.

It is clear that in \eqref{eq:final-sum}, the best we can do is to
make $\tilde \Phi_j$ parallel to the leading eigenvector of $C_{\alpha\beta}$.
{\it This is our answer:} the order parameter operators {\it are} the
eigenvectors of $C_{\alpha\beta}$, and the order-parameter operators
have the same degeneracy as that leading eigenvector. 
The final step is to un-fuse the indices so as to write
each of those eigenvectors as the $D \times D$ matrix
with the matrix elements of the sought-for optimal operator.

At this point, we note that the idea of fusing of indices is
also the essence of a trick used for the Correlation Density Matrix (CDM)~\cite{cheong-CDM}
(see section~\ref{sec:CDM}) but the indices fused here are different.
Thus the relation between the two density matrices
is only mathematical, not physical.

\subsection{Modified viewpoint: singular-value decomposition}
\label{eq:modified_viewpoint}

The focus in the processing in the previous subsection was
to re-mix the {\it state basis} for the cluster.
In contrast, the FMO approach for $M=2$ 
(as well as the modification in Sec.~2.2),
was based on re-mixing the {\it QD indices}.  
In fact, one could have adopted either focus in
either case, so it is not at all obvious whether
these alternate recipes are equivalent.
We shall show that they are, and in the process
will slightly streamline the recipe.

Let us take step back to the QDDM written in fused indices
as $\rho^{\mu}_{\alpha}$. This is a rectangular not a square matrix.
The $\mu$ index has fewer values, so the matrix rank is $M^2-1$ after 
projecting out the average part.

We can perform a singular-value decomposition of the QDDM to obtain
   \beq
      \rho^{\mu}_{\alpha}  = 
              \sum_j \lambda_j V_\mu^{(j)} \tilde{\Phi}_\alpha^{(j)}
   \label{eq:QDDM-SVD}
   \eeq
where $V_\mu$ are the left singular ``vectors'', related to mixing
between QD states, and $\Phi_\alpha$ are the right singular ``vectors'',
relating to mixing of the cluster's Hilbert space.
Thus, the two different focuses used in Sec.~2.2 and in Sec.~3.1 are
just two sides of the same coin: eq.~\eqref{eq:QDDM-SVD} 
unifies the two approaches and shows that the recipes are equivalent. 
The diagonalization in~\eqref{eq:final-sum} 
is nothing more than the trick for converting a singular-value computation into an eigenvalue one. 
The quantity $\lambda_j$ and $\tilde{\Phi}_\alpha^{(j)}$ are identical to those of 
the same name in the previous subsection.
(However, we note that performing the SVD will also yield an operator proportional to $\rhohat_{\av}$.
This operator must be discarded since our desired order parameter is orthogonal to it.)

This, then, is the final version of our recipe: the un-fused version 
of the operators $\tilde{\Phi}_\alpha^{(j)}$ 
associated with the largest singular values $\lambda_j$. 
The reason for the detour through Sec.~3.1 was the need to relate the procedure to 
our objective function $|\sigma_\Phi|^2$.

\subsection{Final processing of order parameters}
\label{sec:Post_process_op}
The output of the QDDM, thus far, is a set of (possibly) degenerate,
orthogonal order parameter operators. This still does not 
solve the problem of identifying the nature of the symmetry-breaking, 
which we now take up. 

The operators extracted must form a representation of the symmetry 
group, and furthermore must form an {\it algebra}. 
(The identity operator will be proportional to the average density matrix,
$\rhohat^\av$.)
Then, by {\it multiplying the operators} with each other 
and decomposing the result into linear combinations of the operator set, 
we {\it obtain this algebra's rules}, computationally.
At this point, inspection should suffice for one to identify which 
of the finite algebras this is. 
Of course, to the extent the operators are derived from a finite system, 
this algebra will not be exact: there will be an error term that is 
not a linear combination of the operators.

We could instead make the operators more transparent by
a pre-preprocessing using lattice or spin symmetries, which have
not been used at all up to this point -- i.e. we got an arbitrary
linear combination of the order parameter operators.
In many cases, we do have a hint for the final processing 
of the order parameter, in the form of the known lattice 
or spin spin symmetries of our Hamiltonian. 
Then distinct components of the order parameter would 
belong to different group representations of those symmetries.

It may happen that the QDDM states will already be nicely
sorted according to those symmetries. In our experience, 
this happens when those symmetries are broken by the choice of the system cell
used for ED, or of the cluster $A$ used to construct the 
reduced density matrices.
But that cannot be relied on: since the splittings of the QD
states are exponentially or algebraically decreasing with system size,
in a sufficiently large system cell the splitting is too small 
to be resolved by Lanczos diagonalization, so the QD eigenstates 
are random linear combinations from the the QD subspace.

If one wanted to force the symmetries onto our states, one has the
choice of doing so at different stages of the construction.
In effect, FMO did this at the earliest state, by rotating
the QD states themselves. Our preference is to impose them
at the end, i.e. on the order parameters themselves, just
prior to computing the operator products so as to tease out
the structure of the algebra that they form.
Indeed, this is more or less the situation for 
both our examples in the following sections, 
where we had to merely confirm our prior suspicion as to 
what symmetries are broken.

\section{First QDDM example: triangular lattice bosons at 1/3 filling}
\label{sec:tV}

We demonstrate the workings of the QDDM for the $t-V$ model of spinless hard core bosons 
on the triangular lattice at 1/3 filling. The Hamiltonian of this model is given by,
\begin{equation}
H = -t \sum_{\langle i,j \rangle} \left( b_i^{\dagger}b_{j} + \text{h.c.} \right) + V \sum_{\langle i,j \rangle} n_i n_j 
    	\label{eq:tV_model}
\end{equation}
$b_i^{\dagger}$ ($b_i$) denotes the particle creation (destruction) operator, 
$n_i = b_i^{\dagger}b_i$ is the number operator on site $i$, $\langle i,j \rangle$ 
refer to nearest neighbor pairs of sites, 
$t>0$ is the hopping and $V>0$ is the repulsion strength. 
In recent years, the phase diagram of this model has been of 
considerable theoretical interest~\cite{tV_triangle,Wessel_tV,Hassan_tV} 
since it supports solid, superfluid and supersolid phases.

\begin{figure}[htpb]
\centering
\includegraphics[width=\linewidth]{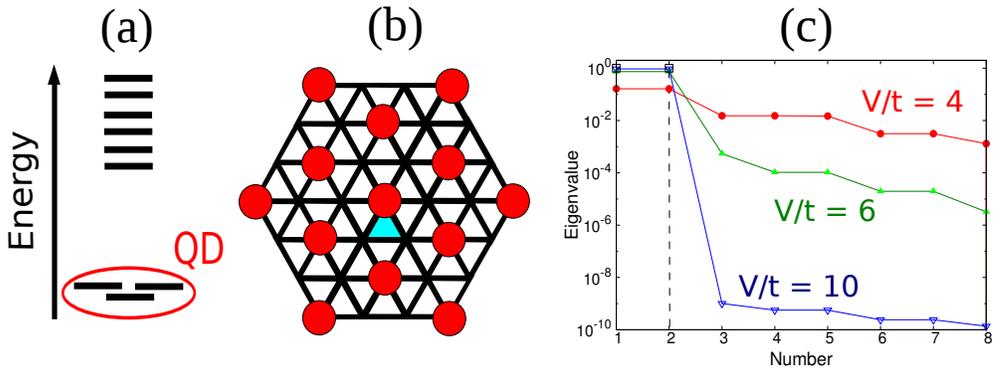}
\caption{(Color online) $t-V$ model 
of spinless bosons at 1/3 filling was considered on various 
unit cells of the triangular lattice. 
(a) Schematic of the low energy spectrum shows three quasidegenerate states (1+2 structure), 
which we use to process the QDDM and the contracted matrix (see Sec~\ref{sec:QDDM}) 
(b) Geometry of the 27 site cell showing the solid phase at 1/3 filling. 
The region (blue/shaded), denoted as $A$ in the text, is the one over which the QDDM is calculated. 
(c) Eigenvalues of the contracted matrix (arranged according to decreasing value) 
at various values of $V/t$ for the 27 site cell. 
Two dominant eigenvalues are observed in the solid phase. (The 
dashed line denotes the structure of eigenvalues in the limit $V/t \rightarrow \infty$.) 
}
\label{fig:tV_triangle} 
\end{figure}	

\begin{figure}[htpb]
\centering
\includegraphics[width=0.8\linewidth]{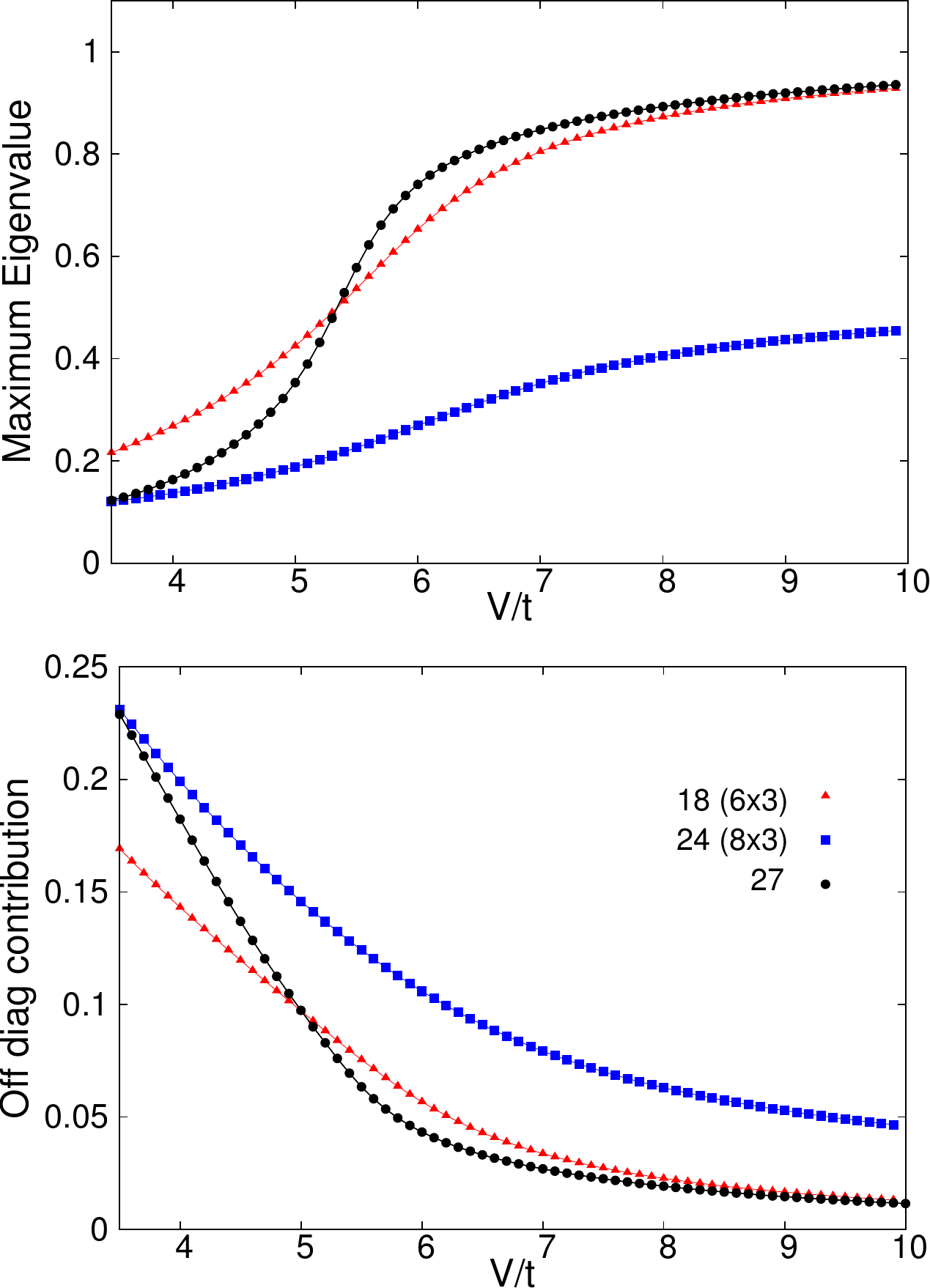}
\caption{(Top) Maximum eigenvalue of the contracted matrix (obtained from the QDDM) as a function of $V/t$ 
corresponding to the dominant operator 
on the elementary triangle. (Bottom) The total off diagonal contributions (sum of squares of 
all off diagonal elements) summed over the first two dominant operators. 
Both metrics display a dependence on the aspect ratio and the total number of sites. 
Based on the observed trends, we expect a genuine phase transition around $V/t \approx 5 - 6$ 
in the thermodynamic limit. }
\label{fig:max_eig_tV} 
\end{figure}	

Our objective here will be to show that the QDDM analysis 
can reproduce the order parameter of the solid phase and signal the 
occurrence of a phase transition as a function of $V/t$. 
To begin with, we observe that at $V/t \rightarrow \infty$, the energy spectrum has exactly 
three degenerate states which become quasidegenerate as soon as $V/t$ is finite 
(schematically depicted in Figure~\ref{fig:tV_triangle}(a)). This QD structure of the 
spectrum persists till approximately $V/t \approx 4$. Thus, $M=3$ QD states are used as an 
input for the QDDM calculation.

\begin{table}[htpb]
\begin{center}
\begin{tabular}{|c||c|c||c|c|}
\hline
Operator & $\Phi_1$ & $\Phi_2$ & $\Phi_1$ & $\Phi_2$  \tabularnewline
\hline
\multicolumn{1}{|c||}{} &\multicolumn{2}{|c||}{$V/t=6$}&\multicolumn{2}{|c|}{$V/t=4$} \tabularnewline
\hline
\multicolumn{5}{|c|}{0 bosons on cluster}\\
\hline
$|000 \rangle \langle 000 |$ & $+0.000$ & $+0.000$ & $+0.000$ & $+0.000$   \tabularnewline
\hline
\multicolumn{5}{|c|}{1 boson on cluster}\\
\hline
$|001 \rangle \langle 001 |$ & $-0.404$ & $-0.699$ & $-0.387$ & $-0.670$   \tabularnewline
\hline                                                                  
$|001 \rangle \langle 010 |$ & $+0.042$ & $-0.073$ & $+0.084$ & $-0.146$  \tabularnewline
\hline                                                                  
$|001 \rangle \langle 100 |$ & $-0.084$ & $+0.000$ & $-0.169$ & $+0.000$   \tabularnewline
\hline                                                                  
$|010 \rangle \langle 010 |$ & $+0.807$ & $+0.000$ & $+0.773$ & $+0.000 $  \tabularnewline
\hline
$|010 \rangle \langle 100 |$ & $+0.042$ & $+0.073$ & $+0.084$ & $+0.146$   \tabularnewline
\hline                                                                  
$|100 \rangle \langle 100 |$ & $-0.404$ & $+0.699$ & $-0.387$ & $+0.670$   \tabularnewline
\hline                                                                  
\multicolumn{5}{|c|}{2 bosons on cluster}\\                 
\hline                                                                  
$|011 \rangle \langle 011 |$ & $+0.010$ & $-0.017$ & $+0.041$ & $-0.072$   \tabularnewline
\hline                                                                  
$|011 \rangle \langle 101 |$ & $-0.004$ & $-0.008$& $-0.022$  & $-0.038$  \tabularnewline
\hline                                                                  
$|011 \rangle \langle 110 |$ & $+0.009$ & $+0.000$& $+0.044$  & $+0.000$  \tabularnewline
\hline                                                                  
$|101 \rangle \langle 101 |$ & $-0.020$ & $+0.000$ & $-0.083$ & $+0.000$  \tabularnewline
\hline                                                                  
$|101 \rangle \langle 110 |$ & $-0.004$ & $+0.008$& $-0.022$  & $+0.038$  \tabularnewline
\hline
$|110 \rangle \langle 110 |$ & $+0.010$ & $+0.017$& $+0.041$  & $+0.072$  \tabularnewline
\hline                                                                  
\multicolumn{5}{|c|}{3 bosons on cluster}\\                 
\hline
$|111 \rangle \langle 111 |$ & $+0.000$ & $+0.000$ & $+0.000$ & $+0.000$   \tabularnewline
\hline
\end{tabular}
\label{tab:ops_V_6}
\caption{Numerical coefficients for the dominant order 
operators in terms of the triangular cluster projection operators for $V/t=6$ and $V/t=4$ 
from the QDDM analysis of the 27 site cell. 
For brevity, the Hermitian conjugate is not shown.} 
\end{center}
\end{table}

We then choose $A$: the region over which the reduced density matrices are computed, 
to be a triangle (see blue (grey) region in Figure~\ref{fig:tV_triangle}(b)). 
Since the dimension of the Hilbert space of $A$ is $D=2^{3}=8$ and 
there are $M=3$ quasidegenerate states, the QDDM $\rho^{mn}_{aa'}$ is 
a $M^2 \times D^2  = 9 \times 64$ matrix and 
the contracted matrix $C_{\alpha\beta}$ is a $D^2 \times D^2 = 64 \times 64$ matrix. 
Due to the symmetries of the problem, two of the three quasidegenerate states 
are actually exactly degenerate. Thus, every exact diagonalization run (with a 
different random seed used to start the Lanczos algorithm) gives different QD eigenstates 
and hence different QDDM matrix elements. However, the 
basis invariant construction of the cost function $\sigma_{\Phi}$ 
guarantees that the contracted matrix $C_{\alpha\beta}$, and hence the 
operators obtained from its singular value decomposition, are always the same in all runs.

For $V/t \rightarrow \infty$ (practically set to $10^{4}$), 
for the 27 site unit cell, we found two dominant operators for
$C_{\alpha\beta}$ with eigenvalues 1 and 1. To interpret 
the corresponding eigenvectors, we un-fuse the $\alpha$ index of the obtained eigenvector $\Phi_{\alpha}$ 
into indices denoting a pair of states of $A$'s Hilbert space, $\alpha \rightarrow aa'$. Thus, 
we recast the $D^2=64$ dimensional eigenvectors into $D \times D = 8 \times 8$ 
Hermitian matrices. Denoting the three sites on the triangle $A$ as 1,2 and 3 and 
the basis states in the occupation number representation by $| n_1 n_2 n_3 \rangle$, 
the two dominant operators are found to be,
\begin{eqnarray}
	\Phi_1 &=& \frac{1}{\sqrt{6}} \left( -| 100 \rangle \langle 100 | + 2 |010 \rangle \langle 010 | - |001\rangle \langle 001| \right) \\
	\Phi_2 &=& \frac{1}{\sqrt{2}} \left( +| 100 \rangle \langle 100 | - |001\rangle \langle 001| \right) 
\label{eq:phi_1_2_V_inf} 
\end{eqnarray} 
(Note that these operators are degenerate and hence one could make other (equally valid) 
linear combinations as well.)

When we lower $V/t$, the contracted matrix still has two dominant eigenvalues, 
but their overall magnitude reduces, as can be seen in Figure~\ref{fig:tV_triangle}(c). 
Simultaneously, the sub-dominant eigenvalues become larger.
(The two dominant eigenvalues are exactly equal on the 27 site cell, 
but not on the 18 ($6 \times 3$) and 24 ($8 \times 3$) site cells. 
This can be attributed to the fact that the three sublattices are not 
perfectly equivalent in the latter cases.) 
In addition, the dominant operators become more off-diagonal 
in the occupation number representation. 
For example, for the 27 site cell, for $V/t=6$ ($V/t=4$), the dominant eigenvalues 
are both $0.74$ ($0.16$) and the first subdominant eigenvalue is $0.00054$ ($0.015$). 
As a numerical example, the two dominant operators for both these cases are summarized in Table 1. 

These notions can be put on a quantitative footing, by tracking 
the evolution of the most dominant eigenvalue as a function 
of $V/t$, as is shown in the top panel of Figure~\ref{fig:max_eig_tV}. 
The relatively rapid change in the magnitude of this dominant eigenvalue around $V/t \approx 5-6$ 
suggests the occurrence of a genuine phase transition. This can also be verified 
by observing how the {\it total} contribution of all off diagonal 
elements of the two most dominant operators changes with $V/t$, as is 
shown in the bottom panel of Figure~\ref{fig:max_eig_tV}. 
(More precisely, this metric was calculated by summing squares of all 
off diagonal coefficients over the two most dominant operators.) 

In summary, our results indicate that the solid melts into another phase (which previous numerical 
studies~\cite{tV_triangle,Wessel_tV,Hassan_tV} indicate to be a superfluid), 
at $(V/t)_c \approx 5-6 $ i.e. $(t/V)_c \approx 0.17 - 0.20 $. 
This compares reasonably well with the Monte Carlo estimate~\cite{Wessel_tV} 
of $ (t/V)_c=0.195 \pm 0.025$ and the DMFT estimate~\cite{Hassan_tV} 
of $(t/V)_c=0.216$. We expect that calculations for bigger 
systems and a thorough understanding of the dependence of the order parameter 
on the aspect ratio of the finite systems studied 
will reveal a more quantitatively accurate picture of this phase transition. 

Finally, we note that a fermionic version involving a 
related $t-V$ model was studied by Motrunich and Lee~\cite{Motrunich_Lee}. 
In this case, the solid at 1/3 filling melts into a renormalized Fermi liquid on lowering $V/t$. 
This phase transition can also be detected by a QDDM analysis~\cite{Changlani_QDDM_MPS}.

\section{Second QDDM example: ``emergent spins'' in randomly diluted antiferromagnets}
\label{sec:perc}
We now present another situation where quasidegenerate states occur 
and where their physical origin is unrelated to spontaneous symmetry breaking. 
Here, the quasidegeneracy will be attributed to the 
formation of (and interaction between) local spin moments 
in the presence of disorder.

\subsection{Previously known results for randomly diluted antiferromagnets}
To understand the context of our chosen example, we briefly summarize 
our research on disordered antiferromagnets~\cite{betheperc,changlani_thesis}. 
These systems are theoretically~\cite{chen,bray-Ali,Sandvik2002} 
and experimentally~\cite{vajk} interesting, 
and unlike their clean counterparts are not as well understood.
While realistic systems involve a host of disorder effects: 
non magnetic impurities, bond disorder, edge effects etc., 
our research primarily focused on the case of dilution of 
a clean system (replacing magnetic ions with non-magnetic ones) 
to the percolation threshold. Percolation~\cite{aharony} provides a 
test bed for investigating the competition between geometrical disorder 
and quantum mechanics. 

\begin{figure}[htpb]
\centering
\includegraphics[width=0.8\linewidth]{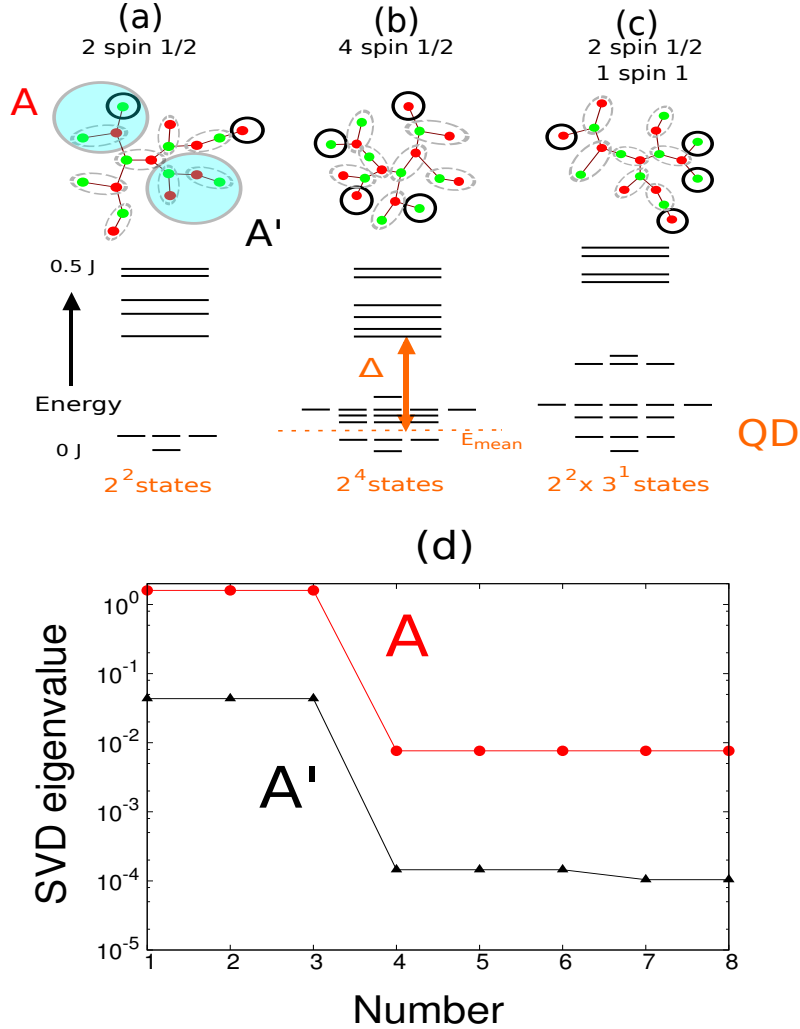}
\caption{(a),(b),(c) show three different percolation clusters (all of the same size) on the Bethe lattice, 
along with their respective low energy spectra. 
The red (dark) and green (light) circles indicate even and odd sublattice sites. The broken dashed lines show dimer coverings 
which serve as a heuristic to locate the ``dangling spins'' (circled with thick black lines). 
Energy spectra for each of the clusters show low-lying quasidegenerate (QD) states separated from the 
continuum by an energy scale $\Delta_{QD}$. For purpose of the QDDM calculation in the text, 
the percolation cluster (a) is chosen. Two different patches $A$ and $A'$ (shown by shaded/blue circles) 
are chosen for two separate QDDM analyses. (d) shows the eigenvalues of the contracted matrix 
calculated for patches $A$ and $A'$.
(Parts of this figure have been adapted from the original by 
Changlani et al.~\cite{betheperc} and modified for the present purposes.)} 
\label{fig:qd_perc} 
\end{figure}

Our work on these systems was motivated by that of 
Wang and Sandvik~\cite{wangandsandvik}, who showed that the 
lowest energy gaps of diluted square lattice Heisenberg antiferromagnets at the percolation 
threshold did not have the same finite size scaling as their clean counterparts 
(whose lowest energy excitations are the usual Anderson ``tower of states''~\cite{Ziman,Gross}). 
They realized the role of regions of ``local imbalance'' (areas on the percolation cluster 
where there is an excess of sites of one sublattice over the other) and they 
devised heuristics for pointing out ``dangling spins'': spins that could 
not be locally paired with another neighboring spin. Figure~\ref{fig:qd_perc}(a,b,c) 
show percolation clusters on the Bethe Lattice where the same qualitative picture of dangling spins 
can be understood. Since the number of such dangling spins at the percolation threshold is 
macroscopically large, their low energy behavior affects 
physical properties such as the low temperature specific heat and susceptibility. 

In a recent paper~\cite{betheperc}, we (in collaboration with others) 
showed the existence of spin 1/2 (and in some cases spin-1) 
``emergent'' degrees of freedom, on randomly diluting the unfrustrated coordination-3 
Bethe lattice Heisenberg antiferromagnet. As is depicted in Figure~\ref{fig:qd_perc}(a-c), 
our inference was based on the observation of quasidegenerate (QD) states seen in the many body spectrum 
obtained from either ED or DMRG. We also found (using multiple metrics), these 
emergent spin-excitations to be relatively localized in space. This picture of the 
low energy physics is very different from the case for clean antiferromagnets 
where the degrees of freedom are macroscopically big sublattice spins.

\subsection{QDDM calculation}
Having established the existence of these local moments, we 
then asked the question: What are the composite ``effective'' spin operators that 
describe the observed QD states? We wish to obtain the answer to this question 
in a relatively unbiased and automated way, for which the QDDM formalism is very apt. 

To demonstrate how the QDDM method works for this system, 
we take as our system the 18-site percolation cluster of 
Figure~\ref{fig:qd_perc}(a), which has two dangling spins 
and hence (we expect) two emergent quasispins, implying a
total of $2^2=4$ QD states~\footnote{We remark that multiple low lying states 
must be targeted in the Lanczos algorithm to ensure that one 
obtains an accurate representation of the QD spectrum. (An insufficient 
number of Krylov vectors may lead to absence of certain QD states.)}. 
Our patch $A$ for the RDM~\footnote{To avoid confusion in using the word ``cluster'' 
for a finite ``percolation cluster'' and the ``cluster $A$'' i.e. our region of interest 
whose RDM is processed, we will instead refer to the latter as a local ``patch'' in this section.} 
is chosen to encompass the central sites participating in one of the emergent spins: namely, three sites forming a 
``fork'', a geometric motif that has a local imbalance associated with it.
The Hilbert space of this patch thus has dimension $2^3$.
\footnote{In fact, as noted in \cite{betheperc}, the emergent spins were found to
be exponentially localized around the fork: thus, by taking a 
larger local patch, we should obtain cleaner numerical results.
The price is not only the exponentially increased computational cost:
if the patch is extended too far, the results may be polluted by
contributions from a different, neighboring emergent spin.}

Processing the QDDM with $M=4$ states, (i.e. one $S=0$ and three lowest $S=1$ states), 
we obtain three dominant operators. 
This is expected because the model is $SU(2)$ symmetric, guaranteeing 
the three operators are those that correspond to the 
three components of an effective spin operator $\bf T$ (basically, 
$T^{z}$, $T^{+}$ and $T^{-}$). 

To be sure of our interpretation, we checked the commutation relations of these operators. 
The operator commutation algebra only {\it approximately} corresponds 
to that of a true spin-$1/2$ algebra, although the approximation is very good in this case. In fact there is apriori no reason that 
the effective spin operator be a {\it true} spin 1/2 operator, since the transformation 
from the bare to effective spins is not a linear/unitary one.
(Also note that we do not see any contributions corresponding 
to the identity operator, because $\rho_{\av}$ is proportional to it. 
Our QDDM formalism explicitly looks for an operator orthogonal to $\rho_{\av}$ 
to split the QD states.)

For the three site fork (site $2$ is the center of the fork and $1$ and $3$ its equivalent ends), 
our numerically obtained operator $T^{+}$ can be expressed 
in terms of the bare spin operators $\bf S_1$, $\bf S_2$ and $\bf S_3$, as 
\beq
	T^{+} \approx 0.085 \left( 2 S_{1}^{+} \left( 1 - 4 S_2 \cdot S_3 \right) + 2 S_{3}^{+} \left( 1 - 4 S_1 \cdot S_2 \right) - S_2^{+} \left( 1 - 4 S_1 \cdot S_3 \right) \right)   
	\label{eq:Tz}
\eeq
(The other operators $T^{-}$ and $T^{z}$ follow from Eq. \eqref{eq:Tz}
since the spin-rotation symmetry is exact at all stages of processing.)

Up to this point we have only discussed one choice of patch 
to process the QDDM. However, one can scan all possible local patches 
and process the QDDM. To show that the QDDM is \emph{not} a misleading 
diagnostic for detecting an effective composite spin operator, 
we considered another patch $A'$ consisting of four sites (shown in Figure~\ref{fig:qd_perc}(a)) 
for the same percolation cluster with two dangling spins. This patch  
does not have any locally unbalanced spins and is expected to be inert 
(i.e. not involved in the low energy physics). As can be seen 
in Figure~\ref{fig:qd_perc}(d), the QDDM analysis of this patch 
revealed that the dominant operator on $A'$ 
has an eigenvalue about $40$ times smaller than that obtained 
for the case of patch $A$. Thus, the QDDM analysis confirms that patch $A'$ 
has no emergent local moments associated with it.

Finally, we remark on the application of QDDM for other percolation clusters. 
For example, the analysis of the 18-sites system in Figure~\ref{fig:qd_perc}(b), should 
be very similar to that already carried out for percolation cluster (a), 
although there will be $M=16$ instead of $M=4$ QD states to process. 
On the other hand, in the case of 
the percolation cluster in Figure~\ref{fig:qd_perc}(c), there are local patches
where the emergent spin 1/2 degrees of freedom 
are on the same sublattice and are located spatially close to one another. 
Thus, the QDDM analysis is expected to reveal the 
effective spin-1 nature of this degree of freedom. 

\section{Extensions of the QDDM method}
\label{sec:extend}
This section collects several ways in which the current
method may be extended: the common factor is they have not
been tried yet, and we offer them in the hopes that others
may take up the challenge. One might extract the low-lying
states by means other than ED (Subsection \ref{sec:Beyond_ED},
or might apply the method to the new systems 
(Subsection \ref{sec:Beyond_SUSY}, and \ref{sec:Beyond_defects}),
in which the quasi-degeneracy or degeneracy has an origin other than
some incipient order.
(One could also imagine modifying the optimization criterion 
or index-fusing methods we used to obtain the results -- 
but we have no specific proposals to offer here.)

\subsection{Beyond Exact Diagonalization: Calculation of the QDDM with Matrix Product States}
\label{sec:Beyond_ED}
We note that the computation of the QDDM 
can be generalized to other numerical methods, 
in particular those that are based on tensor network schemes. 
For example, in the Matrix Product State (MPS) formalism, 
accurate representations for the QD states $|\psi_m \rangle$ can be generated 
by targeting multiple states in the low energy spectrum 
or by performing multiple MPS calculations (i.e. explicitly orthogonalizing an excited state 
MPS with respect to all previously obtained MPSes).

\begin{figure}[htpb]
\centering
\includegraphics[width=0.8\linewidth]{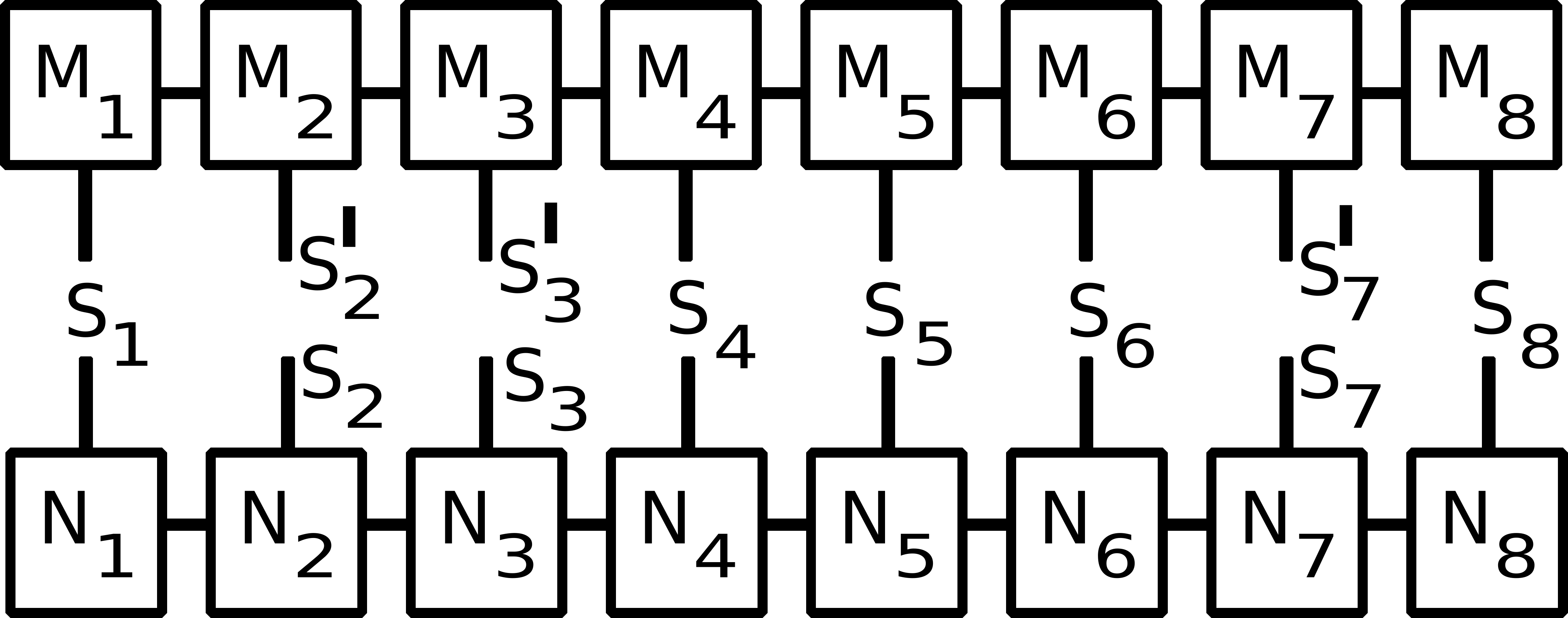}
\caption{Calculation of the QDDM matrix elements using Matrix product states.
MPS corresponding to states $|\psi_m \rangle$ and $| \psi_n \rangle$, denoted by matrices $M_i$ and $N_i$ respectively, 
are obtained from a DMRG calculation. The indices $S_i$ or $S_i'$ are physical indices. 
The region $A$ here comprises of sites 2,3 and 7. Note that its physical indices are {\it not} 
traced over, the rest of the physical indices {\it are} traced over. 
Efficient numerical schemes must be employed to partially contract the above network.} 
\label{fig:cdm_qddm_calc} 
\end{figure}

The QDDM matrix elements between QD states $|\psi_m \ra$ and $|\psi_n \ra$ can be calculated by ``partially contracting'' 
the network formed by their MPS representations, as is depicted diagrammatically 
in Figure~\ref{fig:cdm_qddm_calc}. For illustrative purposes, 
$A$ has been chosen to be a region of 3 sites, numbered 2,3,7, on a 8 site MPS. 
By the term ``partial contraction'', 
we mean that for every pair of states of $A$, trace over all the physical indices 
not belonging to region $A$, in addition to summing over all auxiliary (matrix) indices. 
Note that region $A$ could be a set of local sites in real space (such as the elementary triangle on a triangular lattice), 
but may not remain contiguous when a 2D system is mapped on to a 1D MPS. 
The partial contraction must be carried out computationally efficiently, 
and a scheme for doing so has been discussed by M\"under et al.~\cite{mue10}. 
We are however, unaware of systematic calculations and analyses of~\emph{off-diagonal density matrices} 
obtained from the MPS framework. 
We are exploring this direction and will discuss it elsewhere~\cite{Changlani_QDDM_MPS}.  

\subsection{Fendley-Schoutens supersymmetric model}
\label{sec:Beyond_SUSY}

Fendley and Schoutens~\cite{Fe03,Fe05} introduced a 
remarkable ``supersymmetric'' spinless-fermion lattice model, 
with a Hilbert space restricted so as to disallow 
any nearest-neighbor pairs.  
This model is very favorable to ED as the combined restrictions of 
no spin and no neighbors reduces the Hilbert dimension~\cite{Zh03} 
so the attainable system size is $3-4$ times that for a Hubbard model,
approaching 50 sites (as studied in Ref.~\cite{Gal12}).

The Hamiltonian includes a special potential term, which simply counts the 
number of sites where a fermion could be added without
violating the constraint; roughly speaking, this amounts
to a second-neighbor repulsion and some multi-site interactions.
In addition, the Hamiltonian includes the usual nearest-neighbor hopping 
and chemical potential terms: the supersymmetry holds when all three 
terms have the same coefficients.

From our viewpoint here, the outstanding feature of the model
is that for a range of fillings (about 1/7 to 1/5 in the triangular
lattice case), there is a large-ground state degeneracy~\cite{Gal12,Hui12}
(all of them with exactly zero energy) and the entropy is {\it extensive}. 
To date, there is no field theory for this phase and practically nothing 
is understood about it: is it a like a Hall fluid, an ordered density-wave, 
a Kosterlitz-Thouless critical phase?  

This appears to be an ideal system for applying the QDDM.
(In this case, of course, we have a literal degeneracy rather
than a quasi-degeneracy.)  The dream would be to discover an
algebra of operators which generate all the degenerate 
states, thus allowing us to label all the ground states with 
quantum numbers and providing a framework to ``navigate'' among them.

\subsection{Emergent degrees of freedom, induced by dilution disorder 
in topological or other uniform systems}
\label{sec:Beyond_defects}

Finally, we take inspiration from the example of emergent 
spins on percolation clusters (Sec.~\ref{sec:perc}, above).
Imagine taking a model exhibiting topological order and 
randomly diluting it (or introducing some other form of dilution disorder).
What happens around such a defect has been infrequently studied~\cite{Mi04,Wil10}.

Our interest is in the emergent degree of freedom this may produce,
localized around the defect center. What are the degrees of freedom
of such a ``spin'', and what sort of algebra might they represent?
The naive picture would be that a removed site (or cluster) 
is like a tiny puncture, increasing the system's genus.
Thus, one would expect the same degeneracy due to the topological order, 
as is seen in the large-scale behaviors of the undiluted system 
(presumed known).

But that is not sufficient: the ``puncture'' diameter is comparable
to a lattice constant, so finite-size perturbations spoiling the
topological degeneracy will be major. 
One may imagine an analogy between this emergent ``spin'' and a Heisenberg spin
perturbed (with partial level splitting)
by single-spin anisotropies due to the crystal field. The degenerate states will
split (with some degeneracies remaining) in some pattern which 
depends on the lattice and may not be evident from field theories.
The QDDM, using a cluster $A$ just surrounding the defect region,
appears to be a promising way to find the desired patterns.

After understanding a single degree of freedom, the next step is
to understand the interactions between a {\it pair} of them
(e.g. ~\cite{betheperc,Wil10}). One expects some sort of ``spin''--``spin''
interaction, mediated by the intervening topological fluid.
The Correlation Density matrix (described in the next section)
appears to be well-suited to elucidate such interactions.

An amusing possibility arises if the mediated interactions 
are unfrustrated. This will probably induce {\it long-range order} of the
defect sites at $T=0$. The system might then exhibit a mix of
coexisting topological and long-range orders.

\section{Correlation density matrix}
\label{sec:CDM}

In this section, we turn our attention to another kind of reduced density matrix,
the correlation density matrix (CDM), originally proposed in Ref.~\cite{cheong-CDM}.
The CDM is quite distinct from the QDDM  -- except insofar as
both of these methods are basis-invariant, depend
on density matrices, and may be used to detect 
long-range order when the order parameter is 
not known in advance.  Whereas the QDDM is constructed
using {\it one} cluster and {\it many} eigenstates, 
the CDM uses {\it two} clusters and in principle
{\it one} eigenstate.  Whereas the QDDM detects
{\it global orders} -- either global symmetry-breaking or 
(with an appropriate cluster) topological order, 
the CDM detects {\it correlation functions}. Specifically, the CDM may be used to:
\begin{itemize}
\item
verify the absence of {\it any} kind of order (even exotic
orders that no one has imagined) in systems such as
the spin-1/2 kagom\'e antiferromagnet
\item
verify the existence of long range order (by identifying 
that correlations are not decaying with distance, even
if the nature of the correlation was not recognized in
advance)
\item
extract the scaling exponents and scaling operators of
a critical system, in an extension of the method
\end{itemize}

Critical correlations demand the RDM for a wide range of cluster separations. 
Ground states for the sizes needed are not adequately accessed by Lanczos diagonalizations, 
even for one-dimensional chains. Instead, DMRG may 
be used to obtain very accurate approximations of the ground state and
the CDM can be calculated within the MPS formalism~\cite{mue10}. 
One must, however, be attentive to the usual artifacts of DMRG, such as end effects.

We first outline the basic version of the CDM, 
which analyzes just one particular pair of clusters at 
a time ~\cite{cheong-CDM}. An extension of this 
analysis involves considering all possible separations at the same 
time~\cite{mue10}. Finally, we 
delineate the CDM approach from the superficially similar, 
and more familiar, mutual information approach.

\begin{figure}[htpb]
\centering
\includegraphics[width=0.8\linewidth]{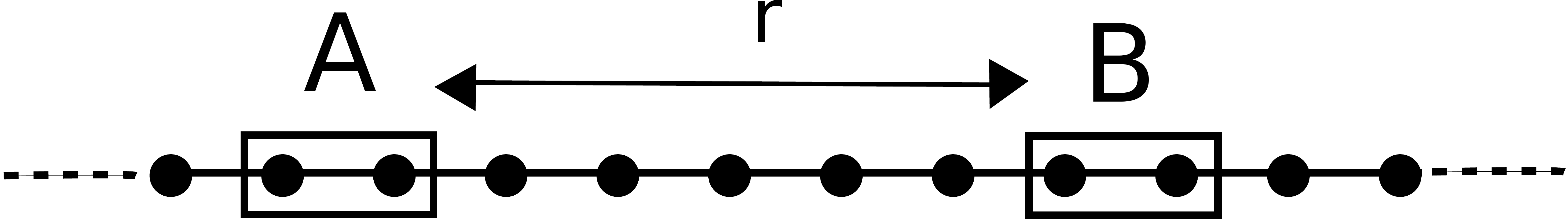}
\caption{The CDM is calculated by computing the density matrix of combination of spatially separated 
regions $A$ and $B$.} 
\label{fig:cdm_geometry} 
\end{figure}	

\subsection{Two-cluster density matrix and correlations}

Take two disjoint clusters $A$ and $B$, for example, as shown in Figure~\ref{fig:cdm_geometry}. 
They should be small -- just large enough
that an exotic order parameter can be defined on them,
e.g. three sites for scalar spin chirality -- and
(normally) congruent to each other.
The aim is to study the correlation between the degrees
of freedom in $A$ and in $B$, as one of them is translated
relative to the other. The CDM ~\cite{cheong-CDM} is defined as 
   \beq
   \hat{\rho}_\corr(\rr) \equiv
    \hat{\rho}_{A\cup B} -\hat{\rho}_A \otimes \hat{\rho}_B,
   \eeq
where $\rhohat_A$ is the density matrix for 
the (small) cluster $A$,
constructed by tracing out its environment, similarly
$\rhohat_B$ and $\rhohat_{A\cup B}$.

Evidently, the CDM contains all possible correlations between 
operator $\Oop_A$ on $A$ and another operator $\Oop_B$ on $B$,
since the connected correlation function is given by
   \beq
     \la \Oop_A \Oop_B \ra_{\rm conn} 
     \equiv \la \Oop_A \Oop_B \ra - \la \Oop_A \ra \la \Oop_B \ra \equiv
     {\rm Tr} (\rhohat_\corr \Oop_A \Oop_B ).
   \label{eq:CDM-all-corr}
   \eeq
Thus, the Frobenius (sum-of-squares) norm $||\rhohat_\corr||$ 
immediately gives an absolute measurement of the total 
correlations of all kinds between the two clusters.~\footnote{
It should be noted that A. L\"auchli and S. Capponi independently
discovered the CDM idea up to this point.}

It may happen that the finite system being diagonalized
has degenerate or quasi-degenerate ground states; this
may simply be a consequence of the specific shape of the
system being diagonalized. Then the recipe~\cite{cheong-2D}
is to restore the symmetry of the Hamiltonian
by averaging over the (quasi) degenerate states 
in the evaluation of $\rhohat_A$, $\rhohat_B$, and 
$\rhohat_{A\cup B}$ (just as was used to evaluate $\rhohat_\av$ in the QDDM case).

\subsection{Fused indices and operator singular-value decomposition}
\label{sec:CDM-fused}

Rather than merely condense the correlations into a single scalar
that quantifies their magnitude, we would like an unbiased way 
to discover what {\it operators} are responsible for this correlation.
This can be done using a special singular-value decomposition,
using a notion of fused indices, analogous to our processing 
of the QDDM but using different indices. 

Let $a$ and $a'$ denote the ``ket'' and ``bra'' indices in 
block $A$, and similarly $b$ and $b'$ in block $B$.
Then replace $(a,a')$ by the fused index $\alpha=\alpha(a,a')$, and 
$(b,b')$ by the fused index $\beta=\beta(b,b')$, write out the 
elements of $\rhohat_\corr$ as a $D^2 \times D^2$ matrix 
with components $r_{\alpha\beta}$.
Perform the singular value decomposition, obtaining
  \beq
        r_{\alpha\beta} = \sum _l \sigma_l 
        \phi^{(l)}_\alpha {\phi'}^{(l)}_\beta.
  \eeq
where $\sigma_l$ are the singular values. When we un-fuse the indices, the left and right 
singular ``vectors'' $\phi^{(l)}_\alpha$ and ${\phi'}^{(l)}_\beta$ 
become operators acting on clusters $A$ and $B$ respectively:
  \beq
        \Phiop^{(l)} = \sum_{aa'} \phi^{(l)}_{aa'} |a\ra\la a'|
  \eeq
  \beq
        \Phiop'{}^{(l)} = \sum_{bb'} \phi'{}^{(l)}_{bb'} |b\ra\la b'|.
  \eeq
Since the singular vectors (either right or left) are mutually
orthogonal, the operators on either cluster are also mutually orthogonal,
according to the Frobenius inner product.

Finally this operator singular-value decomposition yields
   \beq
      \hat{\rho}_\corr
      = \sum _l \sigma_l \Phiop^{(l)}(A) \Phiop'^{(l)}(B),
   \label{eq:CDM-sum}
   \eeq
With this, the density matrix has been decomposed into 
terms representing different kinds of correlation. 
The strengths $\sigma_l$ reveal in an unbiased fashion, 
the respective strengths of these terms;
they control the strength of the output when a
correlation function is computed using \eqref{eq:CDM-all-corr}).

Assuming the system has a unique ground state wavefunction,
this must have the full symmetry of the Hamiltonian and
the symmetries (those ones that act within a cluster)
are carried over to symmetries of $\rhohat_\corr$.
It follows that the terms in \eqref{eq:CDM-sum}
can be classified into symmetry sectors: e.g.
they would be labeled by net spin angular
momentum in a spin-full system, and/or by the change
of particle number (within the block) in a system
where that is conserved globally.
Thus one can distinguish, in an electron system,
whether the superconducting type correlations 
are $s$, $p$, or $d$-wave, whether the 
charge-density-wave-like correlations are 
site-centered or bond-centered, and which of these
classes is dominant for the parameters in question.

\subsection {Scaling operator CDM}
\label{sec:Scaling_operator}

The method, as presented up to now, has the deficiency that 
one requires an independent decomposition~\eqref{eq:CDM-sum} 
be carried out for each possible displacement vector 
$\RR$ between clusters $A$ and $B$.  
Our intuition, however, suggests that only a few operators 
on the cluster that capture all its quantum fluctuations 
have long range effects.  Then we invoke locality -- the distant 
correlation/entanglement  must be propagated via a chain
of mediating sites -- and it follows that the same
operators do this, independent of the location of
the distant cluster.

Therefore, in the CDM analysis, the dominant operators
$\Phihat^{(l)}$ and $\Phihat^{(l')}$  on clusters $A$ and $B$
should be the {\it independent} of the separation vectors 
$\RR$ between $A$ and $B$.
The weight multiplying them is $ \sigma_l \rightarrow \sigma_l(\RR)$; 
it decays (and sometimes oscillates) as a function of $\RR$.  

Our expectation of this algebraic form 
has an explicit theoretical basis, for the special case 
of a {\it critical} state. 
Those characteristic operators $\Phiop^{(l)}$ 
are the {\it scaling operators} emerging from the
renormalization group, each corresponding
to a weight function decaying as a pure
power, $\sigma_l \sim 1/|\RR|^{2x_l}$.
where $x_l$ is defined to be a scaling dimension. 
The scaling extension of the CDM should 
be able (given a sufficient range of separations) 
to resolve several different scaling powers --
at least, those that have different symmetries.

A technical recipe is needed for extracting a set
of local operators $\Phihat^{(l)}$ that simultaneously 
work for all separations, and the functions $\sigma_l(\RR)$. 
Variant recipes can be imagined: one might want to obtain 
the distance-dependent weight $\sigma_l(\RR)$;
but if one {\it knows} the correlations will be algebraic,
one could build that knowledge into the analysis, so as to
extract purer versions of $\Phiop^{l}$.
Sec. 5 of Ref.~\cite{mue10} presents a
recipe to combining the CDM for different separations $\RR$,
in the case of a critical linear chain.  
It involves yet more applications of singular-value decomposition.

Also in Ref.~\cite{mue10}, the method was tested on the model 
of spinless fermions on a two-leg ladder with a nearest-neighbor exclusion. 
For certain limiting values of its parameters, we knew that 
this model was nontrivially equivalent to free fermions~\cite{cheong-CDM,cheong-EGS}, 
and is hence a critical model.  

It appears this method would be well-suited for identifying
the form of the slowest-decaying (but still exponential)
correlations in  the spin-1/2 Kagom\'e antiferromagnet, 
for which ED on systems has been done by numerous authors,
with the current limit at 42 sites~\cite{lau-kag}.
To date, the form of this correlation tail was analyzed only
by calculating the spin-spin correlations and by visual inspections~\cite{lau-kag}.

\subsection{Comparison to the Mutual Information approach}
\label{sec:mutual_info}
Finally, we highlight the difference between using the CDM and the 
more modern approach of using the mutual information~\cite{mutual_info_Alba,
mutual_info_furukawa,mutual_info_Melko,mutual_info_Cirac}, 
which is denoted by $I_{AB}$, and defined to be,
   \beq
      I_{AB} \equiv S(\rho_{A\cup B}) - S(\rho_A)-S(\rho_B)
   \eeq
where $S({\hat\rho})\equiv - {\rm Tr} ({\hat\rho} \ln {\hat\rho})$. 
In addition to being very different mathematical objects (the CDM is a 
matrix, whereas the mutual information is a scalar), 
these are clearly independent quantities. 
The one case where an obvious functional relation exists is
where $\rhohat_A$, $\rhohat_B$, and $\rhohat_{A\cup B}$ all deviate infinitesimally
from the identity matrix,
which physically represents the limit of maximum disorder.
A quick calculation finds the functional relation
$I_{AB} \approx \frac{1}{2} ||\rhohat_\corr||^2$.
(Just use the Taylor expansion of
$S({\hat\rho})$ keeping in mind that ${\rm Tr}(\hat\rho)=1$
for any density matrix.)
However, for the most generic case,
the functional relation between the mutual information and the CDM is not very apparent.

In case of 1D critical lattice systems, one is often 
interested in obtaining {\it multiple} scaling dimensions of the low energy continuum theory. 
The mutual information, for two clusters separated by a finite distance, 
involves in general, a sum over many power laws (corresponding to all 
low lying operators), making it hard to resolve 
the various individual operator contributions to it. While the CDM suffers from similar problems, 
Ref.~\cite{mue10} has presented a scheme taking advantage of the operator nature of the method, 
to process more than one scaling dimension and scaling operator.
It would be interesting to apply this (or related) CDM formalism to more challenging situations, such as critical systems 
with central charge greater than one. We will leave this to future exploration. 
 
\section{Conclusion}
\label{sec:Conclusion}
We now summarize our findings and conclude by briefly mentioning 
some future directions for the RDM based 
methods discussed in this paper.

Our starting point was the appreciation that 
low lying quasidegenerate (QD) states, in addition to the ground state, 
can play an important role in 
the discovery of the system's order parameter. 
The path we followed was one that was proposed 
by Furukawa, Misguich and Oshikawa~\cite{furukawa} (FMO): the method of processing 
{\it their} ``quasidegenerate density matrix'' (QDDM) was reviewed 
in Sec~\ref{sec:FMO} and its limitations were understood. 

In particular, we introduce major improvements to the original QDDM formulation of FMO:
\begin{itemize}
\item[(1)] 
         We clarified (Sec. \ref{sec:Ini_sym_proc})
         the pre-processing of the QD eigenstates,
         showing that prior assumptions about the order, 
         are unnecessary.
\item[(2)] 
         We used $L^2$ norms throughout, 
         in particular replacing the objective function of FMO;
         which opened the door to using the 
         full apparatus of linear algebra,
         in particular the ``operator singular-value decomposition'',
         which depend on the trick of fusing indices.
         (Sec. \ref{sec:FMO-opt-criterion}.)
\item[(3)]
         We extend the method to the fuller general
         case with more than two QD states,
         and symmetry breakings other than the simplest ($Z_2$), 
         kinds (Sec.~\ref{sec:QDDM})
\end{itemize}

The technical trick of {\it fused operators} and
{\it operator singular-value decomposition}~\cite{cheong-CDM} 
appeared more than once. This deals with 
an array of operators (thus a four-index object), 
by combining pairs of indices into one, 
then diagonalizing or constructing a singular-value 
decomposition, and finally un-fusing the indices 
to obtain the result. 
This idea appeared in the QDDM of Sec.~\ref{sec:FMO} (implicitly)
and is central in Sec.~\ref{sec:QDDM},
as well (with a different kind of grouping) for
the ``correlation density matrix" (CDM) of Sec.~\ref{sec:CDM}.
It is hoped that this mathematical machinery will
be useful to define other kinds of DMs and
will simulate research in related directions.

To demonstrate our QDDM, we considered two example systems. 
In Sec.~\ref{sec:tV}, we tested the $t-V$ 
model on the triangular lattice at 1/3 filling where 
we obtained the density wave order parameter and studied 
its behavior on changing $V/t$. The eigenvalue 
corresponding to the order parameter operator 
was found to be a good metric for detecting a 
phase transition in this model. 
We also discussed another case in Sec.~\ref{sec:perc}, 
that was motivated by our studies of randomly diluted antiferromagnets~\cite{betheperc}, where the 
quasidegeneracy arose due to the emergence of 
local spin excitations. We explicitly obtained 
the form of the effective operator 
of a localized spin moment in terms of the 
bare spins on our selected local patches. 

On the technical front, even though 
our current calculations are restricted to 
Lanczos diagonalizations, we discussed 
using the QDDM technique with the DMRG/MPS 
method in Sec.~\ref{sec:Beyond_ED}. 
Ultimately, our hope is to see the application of our QDDM 
method to new problems where a completely unexpected order 
is waiting to be detected. Such examples include 
exactly degenerate zero-energy states in the 
supersymmetric Fendley-Schoutens spinless 
fermion models~\cite{Fe03,Fe05}, discussed in Sec.~\ref{sec:Beyond_SUSY} 
and other problems with dilution disorder, discussed in Sec.~\ref{sec:Beyond_defects}.

Finally, we reviewed the ``correlation density matrix''~(CDM) 
in Section~\ref{sec:CDM} which utilized the ground state 
density matrix of spatially separated regions. 
This DM shares a lot of mathematical apparatus 
in common with the QDDM: however it provides 
complementary information (such as the 
type of long range order in the system or lack there of, 
and the scaling dimensions and operators for critical systems). 
We have also briefly discussed the merits of the CDM with 
respect to approaches which use the mutual information entropy.

\section{Acknowledgement}
We thank Sumiran Pujari and Shivam Ghosh for early 
collaboration on the quasidegenerate density matrix and for 
discussions on various related topics. We acknowledge discussions or
collaborations with 
Siew-Ann Cheong, Stefanos Papanikolau, Jan von Delft, Andreas Weichselbaum, 
Garnet Chan, Shinsei Ryu, Taylor Hughes, Bryan Clark, 
Andreas L\"auchli, Norm Tubman, W. M\"under, Victor Chua and Olabode Sule. 
CLH acknowledges support from NSF DMR-1005466.
HJC was supported by grant DOE FG02-12ER46875. Numerical calculations 
were performed on the Taub campus cluster at UIUC/NCSA.

\end{document}